\documentclass[final,1p,times]{elsarticle}

\usepackage{srcltx}
\usepackage{graphicx}

\usepackage{amssymb}
\usepackage{amsthm}

\begin{document}

\begin{frontmatter}

\title{ The self-consistent determination of HF electroconductivity of strongly
coupled plasmas}
\author[IF]{V.~A.~Sre\'{c}kovi\'{c}}
\author[Ukr]{V.~M.~Adamyan}
\author[IF]{Lj.~M.~Ignjatovi\'{c}}
\author[IF]{A.~A.~Mihajlov}

\address[IF]{Institute of Physics, Pregrevica 118, Zemun,
         Belgrade, Serbia}
\address[Ukr]{Department of Theoretical Physics, Odessa
National University, 65026 Odessa, Ukraine}


\begin{abstract}
Here is presented the calculation of the dynamic electrical
conductivity of fully ionized, strongly coupled plasmas as a
function of the external electric field frequency $\omega$. The
calculations are based on the the formula for the energy-dependent
collision frequency which is determined by means of the Green
function theory methods, as a sum over the Matsubara frequencies.
The domain of extremely high electron density: $10^{21}\leq
n_{e}\leq 10^{24} \textrm{cm}^{-3}$, and for the temperature varying
from $10 \textrm{ kK}$ to $1.000 \textrm{ kK}$ was examined. The
real and imaginary parts of the conductivity for every electron
density are presented in the generalized Drude-like form as a
two-parameter function of the frequency $\omega$ in the region $0 <
\omega < 0.5\omega_{p}$, where $\omega_{p}$ is the plasma frequency.
A good agreement between the obtained results and the existing
theoretical and computing simulation data is shown.

\end{abstract}
\end{frontmatter}

\section{Introduction}

The defining of the dynamic high frequency (HF) electrical
conductivity of the strongly coupled plasma is an actual and very
important problem because such characteristics of plasma as the HF
dielectric function, the coefficients of refractivity and
reflectivity, and skin depth are expressed just in terms of the
dynamic electrical conductivity. However, until recently the
determination of the dynamic plasma conductivity in the wide ranges
of the electron densities $(n_{e})$ and the temperatures $(T)$ was
not taken as the priority task. Namely, in the most cases the
mentioned problem was not investigated as the main one
\cite{kobzev_5,berk2,Morz,Reinh,Kwn}, and  only few authors
presented the results of the direct definition of HF electrical
conductivity \cite{berkovski,Sjg,Fur, berk_A,Reinh_2004} related to
some selected values of $n_{e}$ and $T$.

Because of that, a few years ago a certain research of the HF
conductivity of the plasmas was started. It was thought that the
previous researches of the transport properties of the dense
strongly coupled plasmas free of any external fields
\cite{physleta_91, toploprovodnost}, and the plasmas in the presence
of the constant magnetic field \cite{jphysd_94}, gave some results
which were useful in the case of the plasma in the external time
depending fields. From this reason the research of the HF plasma
conductivity was also devoted to the dense strongly coupled plasmas.
The main aim was to examine the HF conductivity in the whole area of
such plasmas, including here the region of the extremely large
degrees of its non-ideality. This was especially important, since
the existing literature data about the HF plasma conductivity
\cite{berkovski,Sjg,Fur, berk_A,Reinh_2004} are just related to the
extremely strongly coupled plasmas $(10^{21} \textrm{cm}^{-3} <
n_{e} \lesssim 10^{24} \textrm{cm}^{-3})$.

The first results of that research were presented in
\cite{jphysd2001}, where the new method of the calculations of the
strongly coupled plasma HF conductivity is described. This method
was tested on the plasmas with $10^{17} \textrm{cm}^{-3} \leq n_{e}
\leq 10^{19} \textrm{cm}^{-3}$ and $10^{4} \textrm{K} \leq T \leq 5
\cdot 10^{4} \textrm{K}$, in the microwave and far-infrared regions
of $\omega$, were $\omega$ is the frequency of the external electric
field. The developed method uses the energy-dependent electron
collision frequency which is determined as a sum over the Matsubara
frequencies, using the Green function theory methods and the
corresponding self-consistent calculations procedure, which is
described in details in \cite{physleta_91, jphysd_94}. The final
results, i.e. the HF conductivity $\sigma(\omega) \equiv
\sigma(\omega; n_{e}, T)$, as well as its real and imaginary parts
$\sigma_{Re, Im}(\omega) \equiv \sigma_{Re, Im}(\omega; n_{e}, T)$,
were presented in the adequate, three-parameter Drude-like form. As
the second step, by \cite{jphysd2004} the same method with its
improved numerical procedure it was applied to the denser plasmas,
with $10^{19} \textrm{cm}^{-3} \leq n_{e} \leq 10^{21}
\textrm{cm}^{-3}$ and $2\cdot 10^{4} \textrm{K} \leq T \leq 10^{6}
\textrm{K}$, in infrared, visible and near UV regions of $\omega$.

Since, in \cite{jphysd2001, jphysd2004} the domain of low and
moderately non-ideal plasmas was covered, the first aim of this
Letter was to apply the developed method by the adequate numerical
procedure, in order to determine of the HF conductivities of the
strongly and extremely coupled plasmas, i.e. for $10^{21}
\textrm{cm}^{-3} \leq n_{e} \lesssim 10^{24} \textrm{cm}^{-3}$. This
way would not only laboratory strongly coupled plasmas but also some
astrophysical plasmas (for examples, the photospheres of some white
dwarfs \cite{koe80}) be considered.

Another aim is to compare the obtained results with the existing
theoretical results and computer simulation data
\cite{berkovski,Sjg,Fur, berk_A,Reinh_2004}, which refer to $n_{e} =
3.8 \cdot 10^{21} \textrm{cm}^{-3} ,  2.52 \cdot 10^{22}
\textrm{cm}^{-3}$ and $ 1.61 \cdot 10^{24} \textrm{cm}^{-3}$. Such a
comparison is very important because there is no data concerning the
direct experimental measurements of HF conductivity of the dense
strongly coupled plasmas \cite{Reinh}, \cite{Reinh_2000}.

Similarly to our previous papers \cite{jphysd2001,jphysd2004}, the
HF conductivity $\sigma(\omega)$, as well as  $\sigma_{Re,
Im}(\omega)$, are presented here in the three-parameters Drude-like
form, where one of these parameters is the static conductivity
$\sigma_{0}$ of the considered plasma, which is suitable for further
consideration. However, while in \cite{jphysd2001,jphysd2004} the
values of $\sigma_{0}$ could be taken from \cite{jphysd_94}, the
corresponding values for the region $10^{21} \textrm{cm}^{-3} \leq
n_{e} \lesssim 10^{24} \textrm{cm}^{-3}$ were especially calculated
here.

The calculations of $\sigma_{Re}(\omega)$ and $\sigma_{Im}(\omega)$
in all considered cases, for all taken values of $n_{e}$ and $T$,
are performed here in the frequency range $0< \omega < 0.5\cdot
\omega _{p}$, where $ \omega _{p}$ is the plasma electron frequency.
The results are compared with the data from \cite{berkovski,Sjg,Fur,
berk_A,Reinh_2004}.

\section{Model}

\subsection{Basic equations \label{Old RPA}}

In this Letter it is considered a completely ionized plasma in a
homogenous and monochromatic external electric field
$\overrightarrow{E}=\overrightarrow{E_{0}}\cdot exp\{-i\omega t\}$.
According to \cite{jphysd2001}, the dynamic electric conductivity of
a strongly coupled plasma $\sigma (\omega )=\sigma _{Re}(\omega
)+i\sigma_{Im}(\omega)$ is presented by the expressions

\begin{equation}
\sigma (\omega )=\frac{4e^{2}}{3m}\int_{0}^{\infty }\frac{\tau(E)}{
1-i\omega \tau (E)} \left[- \frac{dw(E)}{dE}\right]\rho (E)EdE,
\label{sig_rpa}
\end{equation}

\begin{equation}
\sigma _{Re}(\omega )=\frac{4e^{2}}{3m}\int_{0}^{\infty }\frac{\tau
(E)}{ 1+(\omega \tau (E))^{2}}\left[ -\frac{dw(E)}{dE}\right] \rho
(E)EdE, \label{eq::Re}
\end{equation}

\begin{equation}
\sigma_{Im}(\omega) = \frac{4e^{2}}{3m} \int_{0}^{\infty}\frac{
\omega \tau^2(E)}{1 + (\omega \tau(E))^2}
\left[-\frac{dw(E)}{dE}\right] \rho(E) E dE, \label{eq::Im}
\end{equation}

\noindent where $e$, $m$ and $E$ are the charge, mass and energy of
the free electron, $\tau(E)$ - the relaxation time presented later
in the text, $\rho (E)$ - one-electron states density in the energy
space, $w(E) = [exp(\beta E - \beta \mu) + 1]^{-1}$ - the
Fermi-Dirac distribution function, $\mu$ - the chemical potential of
the ideal gas of the free electrons with the density $n_{e}$ and the
temperature $T$, and $\beta =(k_{B}T)^{-1}$.

Comparing Eq. (\ref{sig_rpa}) by $\omega = 0$ with the expression
for the static conductivity $\sigma_{0}$ of the fully ionized
strongly coupled plasma from \cite{physleta_91,jphysd_94}, one can
see that

\begin{equation}
\sigma(\omega=0) = \sigma_{0} = \frac{4e^{2}}{3m}\int_{0}^{\infty
}\tau(E) \left[- \frac{dw(E)}{dE}\right]\rho (E)EdE.
\label{eq::sig0}
\end{equation}

\noindent Consequently, from Eqs.(\ref{eq::Re}) and (\ref{eq::Im})
it follows the validity of the relations $\sigma _{Re}(\omega = 0) =
\sigma_{0}$, $\sigma_{Im}(\omega = 0) = 0$, which have the
sense of the conditions in the point $\omega = 0$ during the
numerical calculations of $\sigma_{Re}(\omega)$ and
$\sigma_{Im}(\omega)$. The right side of Eq.(\ref{eq::sig0}) has the
Lorentz-like form. Because of that the quantity $\tau (E)$ was
interpreted in \cite{physleta_91,jphysd_94} as the corresponding
electron relaxation time (total electron-ion plus
electron-electron), characterizing the considered plasma in the
absence of the external fields. Since the same quantity $\tau (E)$
is used in the expressions (\ref{sig_rpa}- \ref{eq::Im}) for
$\sigma(\omega)$ just the good understanding of the static
conductivity $\sigma_{0}$ as a function of $n_{e}$ and $T$ has a
very important role in defining the strongly coupled plasma dynamic
conductivity. Because of this fact our research was started in (see
\cite{jphysd2001}) in the region of the plasma electron density
$n_{e} \leq 10^{19} cm^{-3}$, where the basic method of the
determination of the $\sigma_{0}$ from \cite{physleta_91,jphysd_94}
was experimentally verified \cite{rev_e_vitel}.

Accordingly to \cite{physleta_91,jphysd_94}, the relaxation time
$\tau(E)$ is defined in the self-consistent approximation by the
following relations:

\begin{equation}
\begin{array}{l}
\displaystyle{ \tau ^{-1}=\frac{4\pi mn_{e}e^{4}}{\beta
(2mE)^{3/2}}\int \limits_{0}^{\sqrt{ 8mE}/\hbar }\frac{dq}{q}\sum
\limits_{a,\nu }\frac{Z_{a}^{2}\Pi _{a\nu }(q)}{ n_{a}\varepsilon
_{\nu }^{3}(q)}},\\
\displaystyle{\quad \varepsilon _{\nu }=1+\frac{4\pi e^{2}
}{q^{2}}\sum \limits_{a,\nu }Z_{a}^{2}\Pi _{a\nu }}\quad ,
\end{array}
\label{tau}
\end{equation}

\noindent where $a$ is the label for the plasma species, $Z_{a}$ -
the species charge number, $\Pi _{a\nu }(q)$ - corresponding
polarization operator in the random-phase approximation (a simple
loop), $\varepsilon _{\nu }$ - the dielectric function expressed in
terms of these quantities, and the $\nu$ summation is extended over
the Matsubara even frequencies \cite{physleta_91}. Here, the
polarization operator $\Pi _{i(j)\nu }(q)$ of non-degenerate $j$-th
ion component is equal to $\beta n_{i(j)}\delta_{\nu 0}$, and the
RPA polarization operator $\Pi _{e \nu}(q)$ of partially degenerate
free electron component, namely

\begin{equation}\label{pi}
    \Pi_{e \nu }^{RPA}(q) = \frac{m}{q}\left(\frac{k_{F}}{\pi \hbar}\right)^{2}
    \int \limits_{0}^{\infty} \frac{u du}{exp(u^{2}/\theta - \beta \mu) +1}
    ln|\frac{2k_{F}u + q + iQ_{\nu}}{2k_{F}u - q + iQ_{\nu}}|
\end{equation}

\noindent where $k_{F} = (3\pi^{2}n_{e})^{1/3}$, $Q_{\nu} = 4 \pi
\nu m / \hbar^{2} \beta q$, $\nu = 0, \pm 1, ...$, and $\theta =
(2m/\beta)\cdot(\hbar k_{F})^{-2}$.

The main numerical problems, which appear during the calculation of
the static conductivity $\sigma_{0}$, are connected to the defining
of the relaxation time $\tau$. Namely, the number of the members in
the $\nu$ sum in Eq. (\ref{tau}), which has to be calculated
individually, rapidly increases with the increase of the electron
density $n_{e}$. For example, at $n_{e} \sim
10^{24}\textrm{cm}^{-3}$ this number is order of magnitude $\sim
10^{5}$ , while each member itself has to be calculated by a special
procedure. It is clear that the determination of the HF conductivity
$\sigma(\omega)$, requires the volume of the calculations several
times larger than in the case of the determination of $\sigma_{0}$.

\subsection{Parameterized form \label{RPA parame}}

According to \cite{jphysd2001} and \cite{jphysd2004}, the quantities
$\sigma _{Re}$ and $ \sigma _{Im}$ can be presented in the
parameterized form, suitable for the comparison with the
corresponding Drude-like formulas, namely

\begin{equation}  \label{sig_par}
 \sigma_{Re}(\omega) = \sigma_{0} \frac{1}{1 + (\omega
\tau_{0}^{*})^{2}k_{1}^{2}}, \qquad \sigma_{Im}(\omega) = \sigma_{0}
\frac{ (\omega \tau_{0}^{*})k_{2}}{1 + (\omega
\tau_{0}^{*})^{2}k_{1}^{2}},
\end{equation}

\noindent where $\sigma _{0}$ is given by Eq. (\ref{eq::sig0}),
$\tau _{0}^{\ast}$ is the effective relaxation time defined by
$\sigma _{0}=\frac{n_{e}e^{2}}{m}\tau _{0}^{\ast}$ and $k_{1} \equiv
k_{1}(\omega) $ and $k_{2} \equiv k_{2}(\omega)$ are parameters,
which describe the influence of the deviation of the $dw/dE$ from
the $\delta$-function \cite{jphysd_94}. When the electron component
of the considered plasma can be treated as a highly degenerate gas
of the free electrons, both coefficients $k_{1}$ and $k_{2}$ tend to
$1$, and equation (\ref{sig_par}) reduce to the Drude-Lorentz model
\cite{shkarovski}. Using the plasma frequency $\omega _{p}$ and the
dimensionless parameter $f_{0p}$, defined by relations $\omega
_{p}=(4\pi n_{e}e^{2}/m)^{1/2}, f_{0p}= \omega_{p}\tau _{0}^{\ast }
= 4\pi \sigma _{0}/ \omega _{p}$, the parameterized expressions for
$\sigma _{Re}(\omega )$ and $\sigma_{Im}(\omega )$ convert into

\begin{equation}
\sigma _{Re}(\omega )=\sigma _{0} \cdot \frac{1}{1+(\omega /\omega
_{p})^{2}f_{0p}^{2}k_{1}^{2}},\quad \sigma _{Im}(\omega )=\sigma
_{0} \cdot \frac{ (\omega /\omega _{p})f_{0p}k_{2}}{1+(\omega
/\omega _{p})^{2}f_{0p}^{2}k_{1}^{2}}, \label{sig_par_2}
\end{equation}

\noindent Our previous numerical analysis have shown that $k_{1}$
and $k_{2}$ can be cast in the following approximate fitting form

\begin{equation}
k_{j}=k_{j0}-a_{j} \cdot \frac{a_{j}b_{j}(\omega / \omega
_{p})}{1+a_{j}b_{j}(\omega /\omega _{p})},\quad j=1,2, \label{k1 k2}
\end{equation}

\noindent where the adjustment parameters $k_{10},a_{1},b_{1}$ and
$k_{20},a_{2},b_{2}$ are determined numerically from the condition
of the best relation between the 'exact' and fitted values of the
factors $k_{1}$ and $k_{2}$. Let us notify that by just using the
expressions (\ref{sig_par})-(\ref{sig_par_2}) for
$\sigma_{Re}(\omega)$ and $\sigma_{Im}(\omega)$ we get the
possibility to confirm whether Drude-like model (when $k_{1,2}
\approx 1$) has the physical sense for the considered $n_{e}$, $T$
and $\omega$.

\section{Results and discussion}

In this paper it was defined the dynamic conductivity $\sigma
(\omega)$ of completely ionized hydrogen-like plasmas, when in
expression (\ref{tau}) only two species (electron and ion with
$Z=1$) are taken into account. Here we had on mind that the using of
the expressions (\ref{sig_par}) and (\ref{sig_par_2}) for the
determination of $\sigma_{Re}(\omega)$ and $\sigma_{Im}(\omega)$
require the knowledge of the static conductivity $\sigma _{0}$. From
this reason the values of $\sigma _{0}$ are especially determined
here in the considered domain $10^{21}\leq n_{e}\leq
10^{24}\textrm{cm}^{-3}$ and $10\textrm{kK}\leq T\leq
1000\textrm{kK}$. These values are presented in Tables
\ref{tab::sig0_RPA}.

Then, according to \cite{jphysd2001} and \cite{jphysd2004}, the
coefficients $k_{1}$ and $k_{2}$ in Eqs. (\ref{sig_par}) and
(\ref{sig_par_2}) are taken in the form (\ref{k1 k2}), where they
are expressed in terms of the parameters $k_{j0}$, $a_{j}$ and
$b_{j}$, $j=1,2$. The values of these parameters were also computed
in the same domain ($10^{21}\leq n_{e}\leq 10^{24}\textrm{cm}^{-3}$
and $10000\textrm{K}\leq T\leq 1000000\textrm{K}$) and presented in
Table \ref{tab::k1} and Table \ref{tab::k2}.

\begin{figure}[h!]
\centering
      \includegraphics[width=\columnwidth, height=0.75\columnwidth]{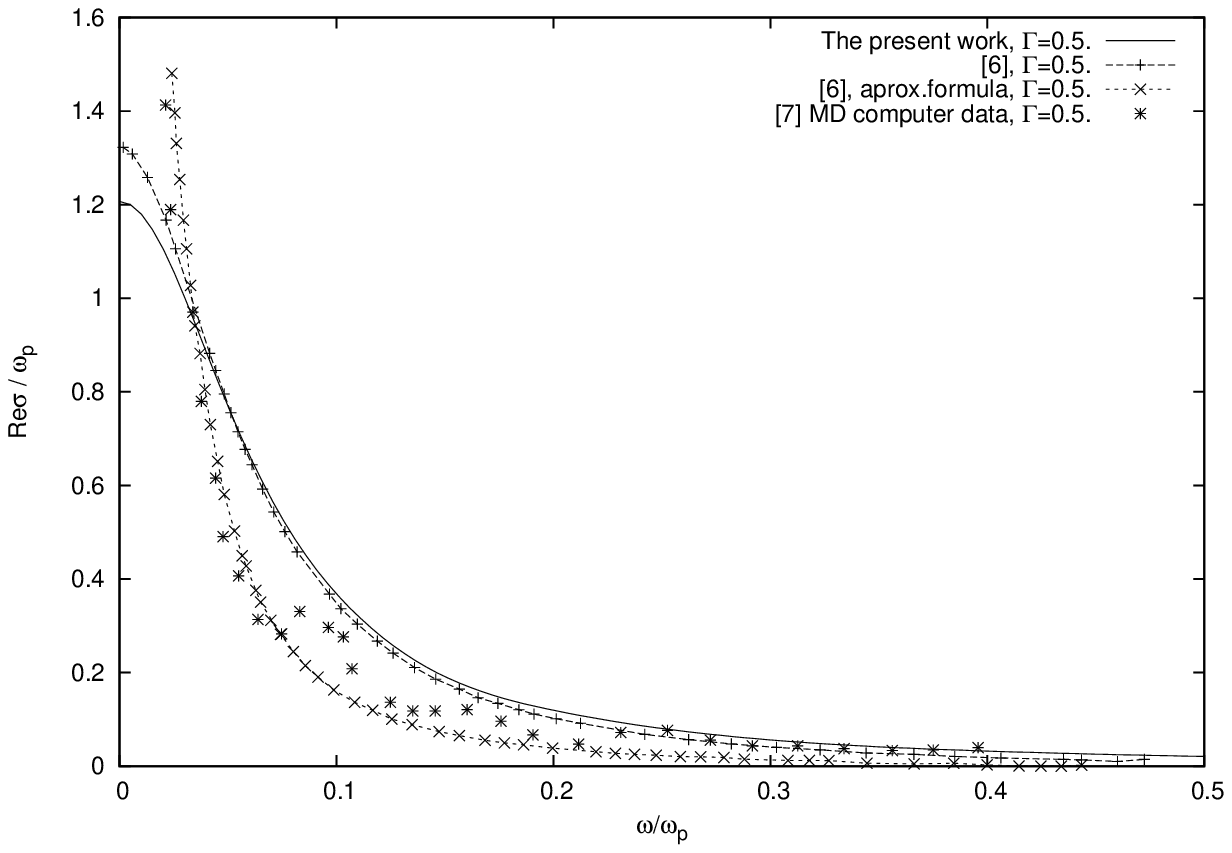}
\caption{ Plasma dynamic conductivity in [CGS] units, as a function
of ($\omega / \omega_{p}$): for $\Gamma = 0.5$ and $r_{s} = 1$, with
the results from \protect \cite{berkovski} and \protect \cite{Sjg},
real part.} \label{fig::berk1a}
\end{figure}

\begin{figure}[h!]
\centering
      \includegraphics[width=\columnwidth, height=0.75\columnwidth]{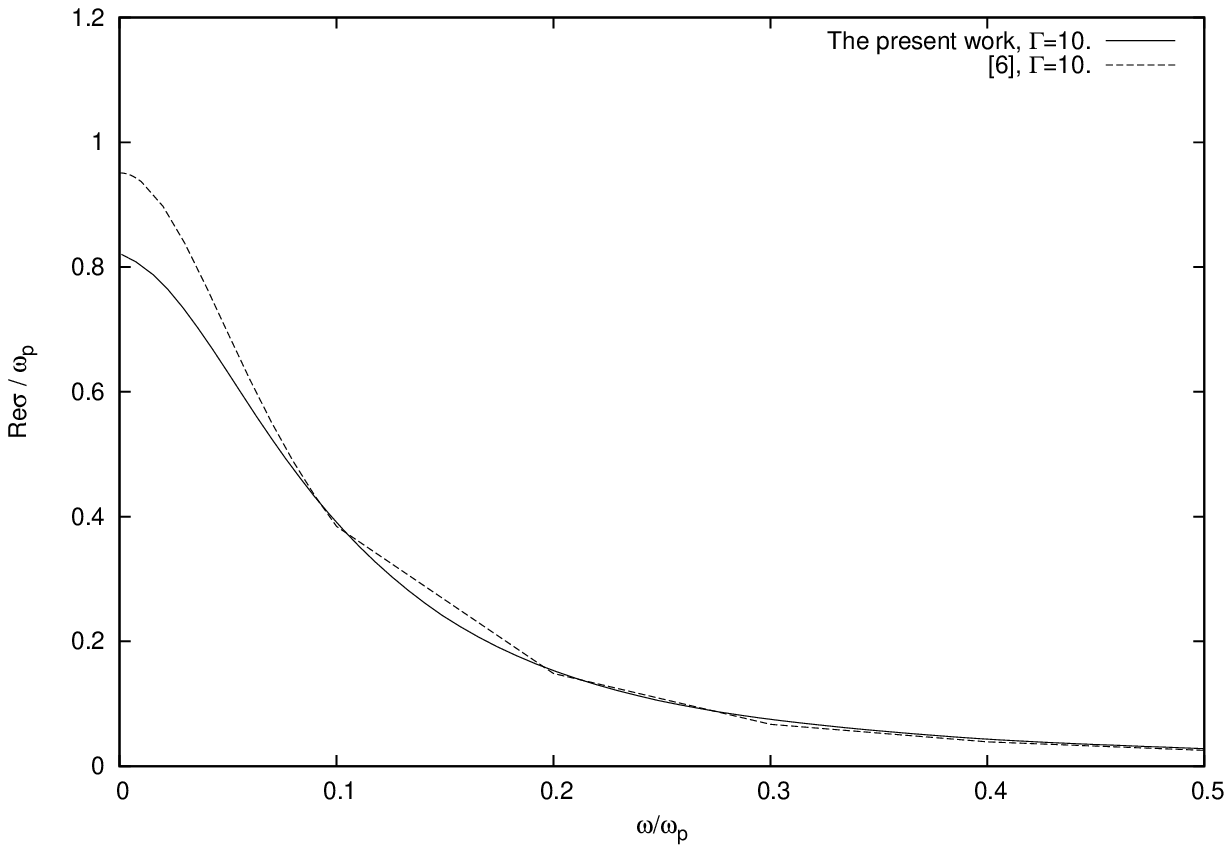}
\caption{ Plasma dynamic conductivity in [CGS] units, as a function
of ($\omega / \omega_{p}$): for $\Gamma = 10$ and $r_{s} = 1$ with
the results from \cite{berk_A}, real part.} \label{fig::berk1c}
\end{figure}

\begin{figure}[h!]
      \centering
      \includegraphics[width=\columnwidth, height=0.75\columnwidth]{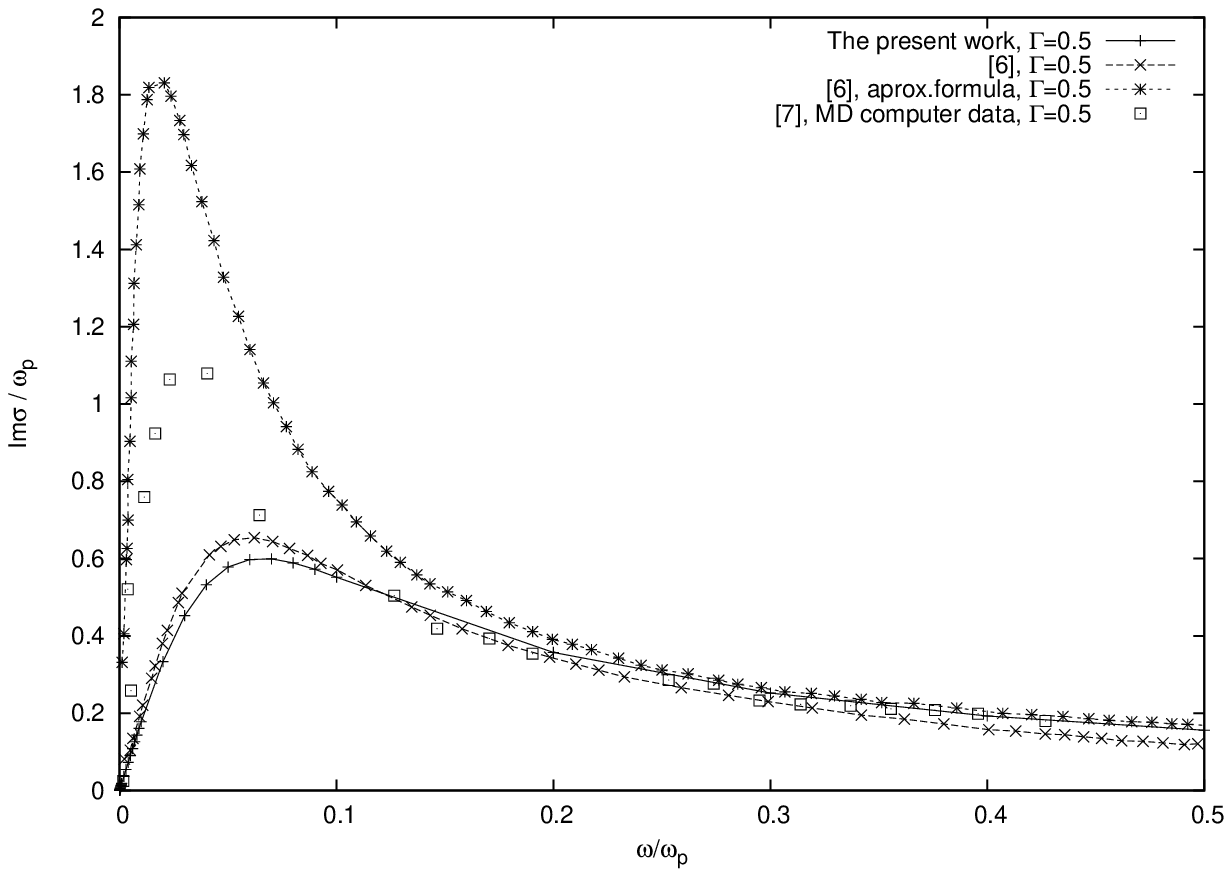}
\caption{ Same as in Fig. \ref{fig::berk1a} but for imaginary
part.} \label{fig::berk1b}
\end{figure}

\begin{figure}[h!]
\centering
      \includegraphics[width=\columnwidth, height=0.75\columnwidth]{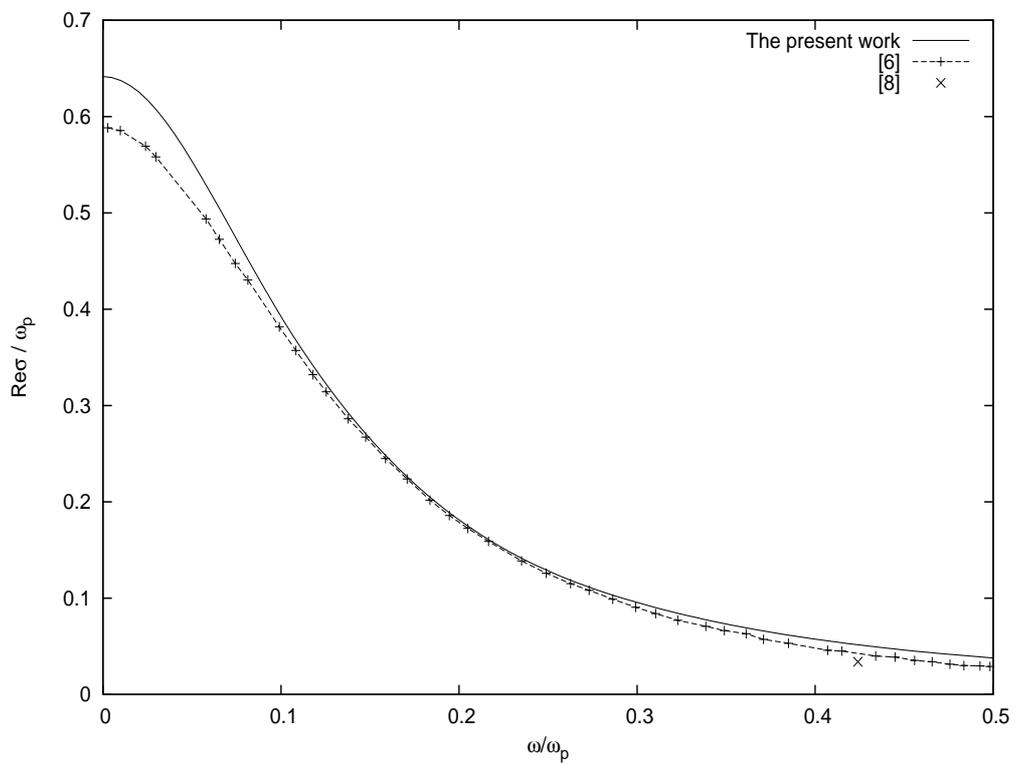}
\caption{ Plasma dynamic conductivity in [CGS] units for $\Gamma =
0.5$ and $r_{s} = 4$, as a function of $(\omega /  \omega_{p})$
frequency ratio, together with results of other authors \protect
\cite{berkovski}  and \protect \cite{Fur}, real part.}
\label{fig::berk2a}
\end{figure}

\begin{figure}[h!]
      \centering
      \includegraphics[width=\columnwidth, height=0.75\columnwidth]{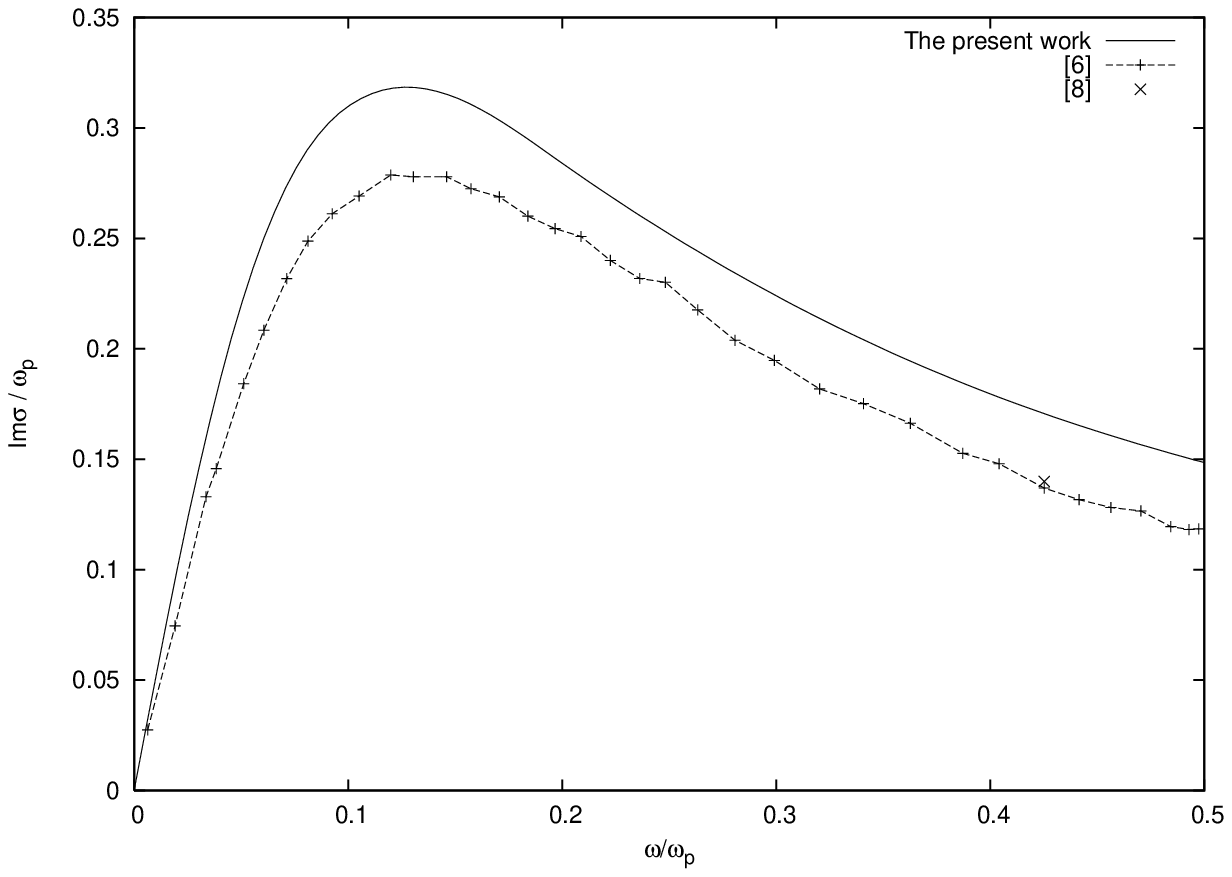}
\caption{ Same as in Fig. \ref{fig::berk2a} but for imaginary
part.} \label{fig::berk2b}
\end{figure}

\begin{figure}[h!]
\centering
      \includegraphics[width=\columnwidth,height=0.75\columnwidth]{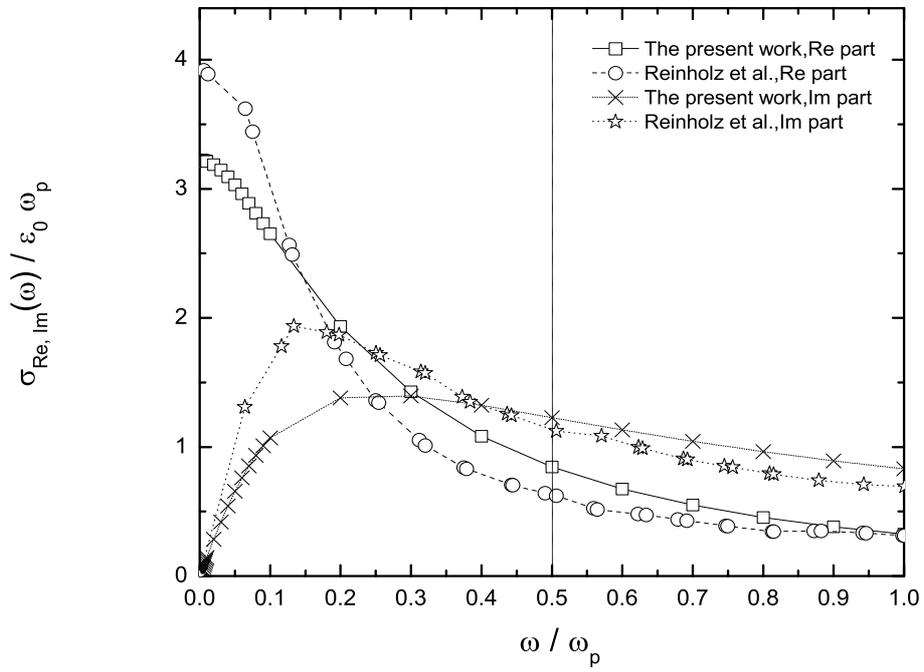}
\caption{Real and imaginary parts of the plasma dynamic conductivity
for $n_{e} = 3.8\cdot 10^{21} \textrm{cm}^{-3}, T = 3.3\cdot 10^4
\textrm{K}$, as a function of $(\omega / \omega_{p})$ frequency
ratio, for $0 \leq \omega / \omega_{p} \leq 1$, together with result
from \protect \cite{Reinh_2004}. The value $\omega / \omega_{p} =
0.5$ especially mark to remind that within our research the region
$0 \leq \omega / \omega_{p} \leq 0.5$ is examined.
\label{fig::reinh}}
\end{figure}

The conduct of $\sigma _{Re}(\omega) $ and $\sigma _{Im}(\omega)$,
calculated directly by Eqs. (\ref{eq::Re}) and (\ref{eq::Im}) for
$10^{21}\leq n_{e}\leq 10^{24}\textrm{cm}^{-3}$ and
$10\textrm{kK}\leq T\leq 1000\textrm{kK}$, is displayed in tables
and figures in the ranges $0<\omega \leq 0.5\cdot \omega _{p}$. The
figures \ref{fig::berk1a}, \ref{fig::berk1c}, \ref{fig::berk2a} and
\ref{fig::reinh}, demonstrate the regular behavior of $\sigma
_{Re}(\omega )$, i.e. the convergence to the corresponding values of
$\sigma _{0}(n_{e},T)$ when $\omega \rightarrow 0$, and the
existence of the interval of variation of $\omega $ where $\sigma
_{Re}(\omega)$ is practically constant. We observe the tendency of
this interval to decrease when temperature $T$ increases. Similarly,
the figures \ref{fig::berk1b}, \ref{fig::berk2b} and
\ref{fig::reinh}, demonstrate a regular behavior of
$\sigma_{Im}(\omega)$, i.e. the convergence to zero when $\omega
\rightarrow 0$, and the presence of a maximum in the interval
$0<\omega <0.5\omega _{p}$. The obtained results, together with the
data from \cite{berkovski}, \cite{berk_A}, \cite{Sjg} and
\cite{Reinh_2004}, show that the position of this maximum comes to
$0.5\omega _{p}$ when $n_{e}$ grows in the region $n_{e}\leq 5\cdot
10^{22}\textrm{cm}^{-3}$, and decrease from $0.5\omega _{p}$ when
$n_{e}$ increases in the region $n_{e}>5\cdot
10^{22}\textrm{cm}^{-3}$.

We compare the behavior of $\sigma _{Re}(\omega)$ and $\sigma
_{Im}(\omega)$, determined in this Letter, with the behavior of the
corresponding quantities determined in: \cite{berkovski} and
\cite{Sjg}, for $\Gamma =0.5$ and $r_{s}=1$; \cite{berkovski},
\cite{Sjg} and \cite{Fur}, for $\Gamma =0.5$ and $r_{s}=4$;
\cite{berk_A} for $\Gamma =10$ and $r_{s}=1$; \cite{Reinh_2004}, for
$n_{e} = 3.8\cdot 10^{21} \textrm{cm}^{-3}$ and $T = 3.3\cdot 10^4
\textrm{K}$. Here $\Gamma =e^{2}/(akT)$ and $r_{s} = a / a_{0}$ are
the well-known non-ideality and Brueckner parameters, where $a_{0}$
and $a=[3/(4\pi n_{e})]^{1/3}$ are the Bohr and the Wigner-Seitz
radii. The results of \cite{Sjg}, \cite{Fur} and \cite{Reinh_2004}
were obtained by using the molecular dynamics (MD) simulation
method, while the results of \cite{berkovski} and \cite{berk_A} are
theoretical. The data from \cite{berkovski} are obtained by two
analytical expressions for the collision frequency (the main one,
obtained within the memory function formalism, and the approximate
one found by the solution of the high-frequency hydrodynamic
equation derived from the macroscopic equations of electron motion).
The corresponding curves in the example $\Gamma =0.5$ and $r_{s}=1$
are shown in Figs.~\ref{fig::berk1a} and \ref{fig::berk1b}, and in
the case $\Gamma =0.5$ and $r_{s}=4$ - in Figs.~\ref{fig::berk2a}
and \ref{fig::berk2b}.

The figures \ref{fig::berk1a} and \ref{fig::berk1b} show that in the
example $\Gamma =0.5$ and  $r_{s}=1$ our values of
$\sigma_{Re}(\omega)$ and $\sigma _{Im}(\omega)$ are mostly in good
relation with the results obtained in \cite{Sjg} by the MD
simulation method, except the region $\omega \lesssim 0.1 \cdot
\omega_{p}$. The same conclusion is valid for the theoretical
results obtained in \cite{berkovski} for $\Gamma =0.5$ and $r_{s}=1$
by means of the mentioned approximate expression. Connected to the
theoretical results from \cite{berkovski} and \cite{berk_A},
obtained by means of the main analytical expression in the cases
$\Gamma =0.5$ and $\Gamma =10$, for the same $r_{s}=1$, a good
agreement with our results exists in the whole region of $\omega$.
Therefore it cold be expected since that main analytical expression
has significant similarity with our expressions.

From figure \ref{fig::berk1b}, concerning $\sigma _{Im}(\omega)$,
one can see that not only our results, but both results of
\cite{berkovski} significantly disagree with the results of the MD
simulation from \cite{Sjg} in the region $\omega \lesssim 0.1 \cdot
\omega_{p}$. We believe that final conclusions, respecting this
fact, can be reached only after some additional MD simulations,
since in this region the behavior of of the static conductivity
$\sigma_{0}$ as a function of $n_{e}$ and $T$ has the dominant role,
and till now we have been considering our values of $\sigma_{0}$ as
fairly reliable.

In the example when $\Gamma =0.5$ and $r_{s}=4$ the figure
\ref{fig::berk2a} shows a good relation of our values of $\sigma
_{Re}(\omega)$ with the values obtained in \cite{berkovski} and
\cite{Fur}, since only in the region $\omega < 0.1 \cdot \omega_{p}$
there exists some appreciable difference (the largest differences
are less that $10$ percents) between our results and those of
\cite{berkovski}. The figure \ref{fig::berk2b}, concerning  $\sigma
_{Im}(\omega)$, shows a qualitative agreement (the differences are
less or close to $20$ percents) of our results with the results of
\cite{berkovski} and \cite{Fur}.

\section{Conclusions}

The results obtained in this Letter, together with the ones presented
in \cite{jphysd2001} and \cite{jphysd2004}, make possible easy
determination of $\sigma _{Re}(\omega)$ and $\sigma _{Im}(\omega)$
in very wide ranges of the electron density and temperature, namely
$10^{17}\leq n_{e}\leq 10^{24}\textrm{cm}^{-3}$ and
$10^{3}\textrm{K}\leq T\leq 10^{6}\textrm{K}$. Consequently, they
can be used for the theoretical research, as well as  for the
interpretation of experimental data in the cases of high pressure
discharge, capillary discharge, shock waves, etc., where the strongly
coupled plasmas, including extremely dense plasmas, are created.
Apart of that, $\sigma _{Re}(\omega)$ and $\sigma _{Im}(\omega)$ can
be useful in some cases of stellar plasmas (e.g., white dwarfs). It
is important that the presented results allow the determination of
other optical characteristics of strongly coupled plasmas: the
permeability, the refractivity, etc. The advantage of the method,
which was used in our research until now, is that it does not
contain any empirical parameters, and have got internal
possibilities for its further improvement.

Let us emphasize that the used procedure is similar to the procedure
for the determination of the conductivity tensor in the case of a
strongly coupled plasma in an external static magnetic field
\cite{jphysd_94}. This fact opens the possibility for the
development of the method of determination of the conductivity of
strongly coupled plasmas in an external high frequency
electromagnetic field in the presence of the static (or
quasi-static) external magnetic field.

Finally, the method used in this Letter does not assume in advance
the quasi-Drude shape for $\sigma(\omega)$, but the way of
presentation of the final expressions for $\sigma_{Re}$ and
$\sigma_{Im}$ automatically demonstrates the character of the
deviations from Drude's expression, at least in the case we
consider.

Certainly, since the numerical procedure for the determination of
$\sigma _{Re}(\omega)$ and $\sigma _{Im}(\omega)$ which is used here
is quite complicated, it is desirable to develop some other methods
which would allow an easier calculation of these quantities. Apart
of that, in the lower part of the considered domain of the electron
density ($n_{e}\leq 10^{20}\textrm{cm}^{-3}$) it is necessary to
develop the method of determination of the dynamic conductivity
which would take into account the presence of the plasma neutral
component.

\section{Acknowledgments}

The presented work is performed within the Project 141033 "Non-ideal
laboratory and ionosphere plasma : properties and applications"
financed by the Ministry of Science of the Republic of Serbia, as
well as the INTAS (GSI-INTAS Project 06-1000012-8707).

\bibliographystyle{elsarticle-num}


\begin{table}[\columnwidth]
\caption{The computed static plasma conductivity $\protect \sigma
_{0}$ as a function of the electron density $n_{e}$ and temperature
$T$, $\left[ 10^{3}(\Omega \textrm{m})^{-1}\right] $.}
\label{tab::sig0_RPA}
\begin{center}
\begin{tabular}{@{}cccccccc}
\hline \hline
\multicolumn{1}{c}{$T$ $[K]$} & \multicolumn{7}{c}{$n_e$ $[\textrm{cm}^{-3}]$} \\
\cline {2-8} & \multicolumn{1}{c}{$10^{21}$} & \multicolumn{1}{c}{$5
\cdot 10^{21}$} & \multicolumn{1}{c}{$10^{22}$} &
\multicolumn{1}{c}{$5 \cdot 10^{22}$} &
\multicolumn{1}{c}{$10^{23}$} & \multicolumn{1}{c}{$5 \cdot
10^{23}$} & \multicolumn{1}{c}{$10^{24}$} \\ \hline
10000        & $ 28.56$  & $ 78.23$  & $ 122.1$  & $ 388.0$  & $ 677.7$  & $ 2579$   & $ 4016$ \\
15000        & $ 33.90$  & $ 86.68$  & $ 135.7$  & $ 410.4$  & $ 701.0$  & $ 2698$   & $ 5175$ \\
20000        & $ 40.17$  & $ 93.34$  & $ 145.2$  & $ 430.8$  & $ 723.2$  & $ 2721$   & $ 4886$ \\
30000        & $ 54.19$  & $ 108.2$  & $ 160.8$  & $ 464.6$  & $ 762.6$  & $ 2767$   & $ 5037$ \\
50000        & $ 85.37$  & $ 145.7$  & $ 198.4$  & $ 511.8$  & $ 824.2$  & $ 2816$   & $ 5070$ \\
100000       & $ 174.2$  & $ 257.6$  & $ 320.1$  & $ 636.6$  & $ 948.7$  & $ 2920$   & $ 5070$ \\
200000       & $ 382.8$  & $ 516.8$  & $ 606.6$  & $ 985.9$  & $ 1305$   & $ 3200$   & $ 5228$ \\
500000       & $ 1165$ & $ 1461$   & $ 1639$   & $ 2278$   & $ 2726$   & $ 4779$   & $ 6669$ \\
1000000      & $ 2806$ & $ 3393$   & $ 3727$   & $ 4829$   & $5531$
& $ 8282$    & $ 10430$ \\ \hline
\end{tabular}
\end{center}
\end{table}

\begin{table}[\columnwidth]
 \caption{Parameters $k_{10}$ (first row for each $T$),
$a_1$ and $b_1$ (second and third rows, respectively) as functions
of $n_e$ and $T$.} \label{tab::k1}
\begin{center}
\begin{tabular}{@{}cccccccc}
\hline \hline
\multicolumn{1}{c}{$T$ $[K]$} & \multicolumn{7}{c}{$n_e$ $[\textrm{cm}^{-3}]$} \\
\cline {2-8} & \multicolumn{1}{c}{$10^{21}$} & \multicolumn{1}{c}{$5
\cdot 10^{21}$} & \multicolumn{1}{c}{$10^{22}$} &
\multicolumn{1}{c}{$5 \cdot 10^{22}$} &
\multicolumn{1}{c}{$10^{23}$} & \multicolumn{1}{c}{$5 \cdot
10^{23}$} & \multicolumn{1}{c}{$10^{24}$} \\ \hline
10000    & $ 1.36558$    & $ 1.14298$    & $ 1.11647$    & $ 1.04503$    & $ 1.02252$    & $ 1.03775$    & $ 1.23393$ \\
         & $ 0.91630$    & $ 0.42670$    & $ 0.36160$    & $ 0.11570$    & $ 0.05374$    & $ 0.00832$    & $ 0.00572$ \\
         & $ 0.80560$    & $ 1.46304$    & $ 1.71715$    & $ 9.66853$    & $ 29.2500$    & $ 447.500$    & $ 975.700$ \\
15000    & $ 1.56925$    & $ 1.21020$    & $ 1.14482$    & $ 1.07335$    & $ 1.04339$    & $ 1.00815$    & $ 0.96590$ \\
         & $ 1.12789$    & $ 0.47940$    & $ 0.35200$    & $ 0.17550$    & $ 0.09703$    & $ 0.01560$    & $ 0.00541$ \\
         & $ 1.06607$    & $ 2.22226$    & $ 2.84164$    & $ 7.27586$    & $ 17.5200$    & $ 244.600$    & $ 1032.00$ \\
20000    & $ 1.70187$    & $ 1.31099$    & $ 1.19312$    & $ 1.09268$    & $ 1.06185$    & $ 1.01567$    & $ 1.02906$ \\
         & $ 1.22186$    & $ 0.63100$    & $ 0.40760$    & $ 0.20970$    & $ 0.13380$    & $ 0.02620$    & $ 0.01223$ \\
         & $ 1.40904$    & $ 2.20497$    & $ 3.36852$    & $ 6.89132$    & $ 13.5600$    & $ 147.200$    & $ 459.700$ \\
30000    & $ 1.85027$    & $ 1.50720$    & $ 1.33432$    & $ 1.12245$    & $ 1.08882$    & $ 1.02599$    & $ 1.01296$ \\
         & $ 1.30102$    & $ 0.90070$    & $ 0.61770$    & $ 0.24750$    & $ 0.18120$    & $ 0.04943$    & $ 0.02285$ \\
         & $ 2.22529$    & $ 2.19804$    & $ 3.11968$    & $ 7.66151$    & $ 11.4900$    & $ 80.3700$    & $ 252.400$ \\
50000    & $ 1.98074$    & $ 1.74157$    & $ 1.58508$    & $ 1.21653$    & $ 1.13706$    & $ 1.05221$    & $ 1.02846$ \\
         & $ 1.36853$    & $ 1.13885$    & $ 0.94640$    & $ 0.37520$    & $ 0.24640$    & $ 0.09391$    & $ 0.05013$ \\
         & $ 4.09404$    & $ 3.00424$    & $ 3.14523$    & $ 7.67889$    & $ 11.7600$    & $ 44.5000$    & $ 117.400$ \\
100000   & $ 2.09206$    & $ 1.95358$    & $ 1.85948$    & $ 1.51913$    & $ 1.34490$    & $ 1.10857$    & $ 1.06780$ \\
         & $ 1.45707$    & $ 1.31413$    & $ 1.21608$    & $ 0.79270$    & $ 0.53610$    & $ 0.17830$    & $ 0.11230$ \\
         & $ 9.00543$    & $ 5.95428$    & $ 5.25356$    & $ 5.95785$    & $ 8.79760$    & $ 29.6900$    & $ 56.2000$ \\
200000   & $ 2.16088$    & $ 2.07644$    & $ 2.02196$    & $ 1.81532$    & $ 1.67551$    & $ 1.28276$    & $ 1.16478$ \\
         & $ 1.53680$    & $ 1.43764$    & $ 1.37555$    & $ 1.13930$    & $ 0.96530$    & $ 0.41640$    & $ 0.24720$ \\
         & $ 19.0300$    & $ 12.0400$    & $ 10.2000$    & $ 7.99801$    & $ 8.30448$    & $ 19.0000$    & $ 34.4800$ \\
500000   & $ 2.21152$    & $ 2.16940$    & $ 2.13967$    & $ 2.03771$    & $ 1.96910$    & $ 1.70479$    & $ 1.53717$ \\
         & $ 1.59191$    & $ 1.54523$    & $ 1.51208$    & $ 1.39149$    & $ 1.30924$    & $ 0.97700$    & $ 0.75090$ \\
         & $ 57.9900$    & $ 32.3600$    & $ 25.8200$    & $ 16.9600$    & $ 14.9600$    & $ 14.5800$    & $ 18.2200$ \\
1000000  & $ 2.26912$    & $ 2.20725$    & $ 2.19266$    & $ 2.12778$    & $ 2.08724$    & $ 1.93392$    & $ 1.82624$ \\
         & $ 1.65679$    & $ 1.58655$    & $ 1.56883$    & $ 1.49589$    & $ 1.44904$    & $ 1.26324$    & $ 1.12668$ \\
         & $ 138.900$    & $ 76.0200$    & $ 59.2600$    & $ 34.3800$    & $ 28.2200$    & $ 20.7800$    & $ 20.2900$ \\ \hline
\end{tabular}
\end{center}
\end{table}

\begin{table}[\columnwidth]
\caption{Parameters $k_{20}$ (first row for each $T$), $a_2$ and
$b_2$ (second and third rows, respectively) as functions of $n_e$
and $T$.} \label{tab::k2}
\begin{center}
\begin{tabular}{@{}cccccccc}
\hline \hline
\multicolumn{1}{c}{$T$ $[K]$} & \multicolumn{7}{c}{$n_e$ $[\textrm{cm}^{-3}]$} \\
\cline {2-8} & \multicolumn{1}{c}{$10^{21}$} & \multicolumn{1}{c}{$5
\cdot 10^{21}$} & \multicolumn{1}{c}{$10^{22}$} &
\multicolumn{1}{c}{$5 \cdot 10^{22}$} &
\multicolumn{1}{c}{$10^{23}$} & \multicolumn{1}{c}{$5 \cdot
10^{23}$} & \multicolumn{1}{c}{$10^{24}$} \\ \hline
10000    & $ 1.22578$    & $ 1.09158$    & $ 1.07647$    & $ 1.02936$    & $ 1.01445$    & $ 1.03622$    & $ 1.23283$ \\
         & $ 0.85360$    & $ 0.43490$    & $ 0.38030$    & $ 0.11700$    & $ 0.05381$    & $ 0.00821$    & $ 0.005572$ \\
         & $ 0.73820$    & $ 1.23034$    & $ 1.42388$    & $ 9.08148$    & $ 28.4700$    & $ 451.400$    & $ 1033$ \\
15000    & $ 1.36175$    & $ 1.12968$    & $ 1.09216$    & $ 1.04821$    & $ 1.02833$    & $ 1.00524$    & $ 0.9649$ \\
         & $ 1.05294$    & $ 0.44090$    & $ 0.34460$    & $ 0.17850$    & $ 0.09769$    & $ 0.01544$    & $ 0.005332$ \\
         & $ 0.96230$    & $ 2.16439$    & $ 2.57804$    & $ 6.67217$    & $ 16.6700$    & $ 246.000$    & $ 1066$ \\
20000    & $ 1.45309$    & $ 1.19154$    & $ 1.11970$    & $ 1.06089$    & $ 1.04061$    & $ 1.01077$    & $ 1.02666$ \\
         & $ 1.13563$    & $ 0.56690$    & $ 0.37520$    & $ 0.21280$    & $ 0.13520$    & $ 0.02613$    & $ 0.01221$ \\
         & $ 1.26874$    & $ 2.20112$    & $ 3.31066$    & $ 6.26462$    & $ 12.6900$    & $ 145.600$    & $ 457.6$ \\
30000    & $ 1.55543$    & $ 1.32016$    & $ 1.20668$    & $ 1.07872$    & $ 1.05821$    & $ 1.01668$    & $ 1.0084$ \\
         & $ 1.19391$    & $ 0.81940$    & $ 0.55200$    & $ 0.24180$    & $ 0.18220$    & $ 0.04959$    & $ 0.02282$ \\
         & $ 2.01639$    & $ 2.10622$    & $ 3.14718$    & $ 7.16940$    & $ 10.6200$    & $ 77.8800$    & $ 249.7$ \\
50000    & $ 1.64423$    & $ 1.48078$    & $ 1.37339$    & $ 1.13410$    & $ 1.08691$    & $ 1.03422$    & $ 1.01839$ \\
         & $ 1.23540$    & $ 1.04367$    & $ 0.86190$    & $ 0.33860$    & $ 0.23370$    & $ 0.09454$    & $ 0.05031$ \\
         & $ 3.76200$    & $ 2.76140$    & $ 2.98249$    & $ 7.80417$    & $ 11.4000$    & $ 42.0900$    & $ 113.6$ \\
100000   & $ 1.71861$    & $ 1.62588$    & $ 1.56188$    & $ 1.32945$    & $ 1.21494$    & $ 1.06960$    & $ 1.04423$ \\
         & $ 1.29643$    & $ 1.18484$    & $ 1.10384$    & $ 0.71430$    & $ 0.47780$    & $ 0.17160$    & $ 0.1116$ \\
         & $ 8.46413$    & $ 5.51045$    & $ 4.84527$    & $ 5.80599$    & $ 8.93956$    & $ 28.6500$    & $ 53.33$ \\
200000   & $ 1.76372$    & $ 1.70824$    & $ 1.67192$    & $ 1.53184$    & $ 1.43625$    & $ 1.17662$    & $ 1.10373$ \\
         & $ 1.36115$    & $ 1.28090$    & $ 1.23103$    & $ 1.03249$    & $ 0.87480$    & $ 0.37330$    & $ 0.2281$ \\
         & $ 17.9300$    & $ 11.3300$    & $ 9.55919$    & $ 7.45556$    & $ 7.86760$    & $ 19.4000$    & $ 34.65$ \\
500000   & $ 1.79739$    & $ 1.76921$    & $ 1.74994$    & $ 1.68242$    & $ 1.63638$    & $ 1.45691$    & $ 1.34383$ \\
         & $ 1.40905$    & $ 1.37086$    & $ 1.34419$    & $ 1.24460$    & $ 1.17609$    & $ 0.88420$    & $ 0.677$ \\
         & $ 53.6500$    & $ 30.0600$    & $ 24.1100$    & $ 15.9400$    & $ 14.0400$    & $ 13.9000$    & $ 17.87$ \\
1000000  & $ 1.82998$    & $ 1.79479$    & $ 1.78467$    & $ 1.74212$    & $ 1.71534$    & $ 1.61265$    & $ 1.53976$ \\
         & $ 1.45044$    & $ 1.40480$    & $ 1.39017$    & $ 1.33254$    & $ 1.29409$    & $ 1.13771$    & $ 1.01851$ \\
         & $ 130.200$    & $ 70.5400$    & $ 54.8300$    & $ 31.8700$    & $ 26.2700$    & $ 19.5000$    & $ 19.17$ \\ \hline
\end{tabular}
\end{center}
\end{table}

\end{document}